\documentclass[12pt,preprint]{aastex}
\usepackage{amssymb}
\usepackage{epsfig}

\newcommand{\etal}{et~al.~}

\begin{document}


\title{Optical Line Emission in Brightest Cluster Galaxies at 0 $<$ \lowercase{z} $<$ 0.6: Evidence for a Lack of Strong Cool Cores 3.5 Gyr Ago?}


\author{Michael McDonald$^{1,2}$}
\altaffiltext{1}{Kavli Institute for Astrophysics and Space Research, MIT, Cambridge, MA 02139, USA}
\altaffiltext{2}{Email: mcdonald@space.mit.edu}


\begin{abstract}
In recent years the number of known galaxy clusters beyond $z\gtrsim0.2$ has increased drastically, with the release of multiple catalogs containing $>$30,000 optically-detected galaxy clusters over the range $0<z<0.6$. Combining these catalogs with the availability of optical spectroscopy of the brightest cluster galaxy from the Sloan Digital Sky Survey allows for the evolution of optical emission-line nebulae in cluster cores to be quantified. For the first time, the continuous evolution of optical line emission in brightest cluster galaxies over the range $0<z<0.6$ is determined. A minimum in the fraction of BCGs with optical line emission is found at $z\sim0.3$, suggesting that complex, filamentary emission in systems such as Perseus A are a recent phenomenon. Evidence for an upturn in the number of strongly-emitting systems is reported beyond $z>0.3$, hinting at an earlier epoch of strong cooling. We compare the evolution of emission line nebulae to the X-ray-derived cool core fraction from the literature over the same redshift range and find overall agreement, with the exception that an upturn in the strong CC fraction is not observed at $z>0.3$. The overall agreement between the evolution of cool cores and optical line emission at low redshift suggests that emission-line surveys of galaxy clusters may provide an efficient method of indirectly probing the evolution of cool cores and, thus, provide insights into the balance of heating and cooling processes at early cosmic times.
\end{abstract}




\section{Introduction}

The high densities and low temperature of the intracluster medium (hereafter ICM) in the cores of some galaxy clusters, dubbed ``cool cores'' (hereafter CCs) due to the steep rise of the density profile and drop in the temperature profile, imply massive amounts of gas cooling radiatively out of the ICM. Early studies \citep[see review by][]{fabian94} suggested that cooling flows should deposit on the order of 100-1000 M$_{\odot}$ yr$^{-1}$ of cold gas onto the brightest cluster galaxy (hereafter BCG) in the cluster core. The lack of evidence for such vast amounts of cold molecular gas and young stars is often referred to as the ``cooling flow problem''.  In recent years, the understanding of cooling flows has benefited immensely from several multi-wavelength observing campaigns, which suggest that $\sim$90\% of the cooling is halted by some form of feedback, while a smaller but still significant supply of cooling gas makes its way to the cluster core.

The presence of optical emission-line nebulae has been noted in a large number of CC clusters to date, the most remarkable being Perseus A \citep{cowie80, conselice01, hatch06}. While the ionization mechanism of these filaments remains a topic of rich debate, the cool gas itself most likely originated in the hot ICM \citep[e.g., ][]{heckman89, crawford05, mcdonald10} and is either undergoing initial infall or has been uplifted from the core either by a buoyant bubble \citep[e.g.,][]{churazov01} or a radio jet \citep[e.g.,][]{werner11}. An important aspect of these nebulae which has received little attention is their evolution over time. The fraction of galaxy clusters with optically-emitting BCGs as a function of redshift may provide valuable insights into the relative effectiveness of heating and cooling processes in the ICM over cosmic time, which is critical for models of galaxy cluster and structure formation.. Furthermore, the fact that CCs tend to have excess optical line emission \citep[e.g.,][]{cavagnolo09, mcdonald10} suggests that tracing the evolution of optically-emitting BCGs may indirectly probe the evolution of CCs in galaxy clusters, which proves exceedingly difficult using X-ray techniques at high redshift.

The purpose of this Letter is to examine the evolution of optical emission-line nebulae in galaxy clusters as a function of redshift. In \S2, the data and analysis techniques are presented, with a focus on aperture corrections. The evolution of optical line-emitting BCGs is presented in \S3 over the range $0<z<0.6$ and the implications of this evolution are discussed. In \S4 we compare this evolution to that of the CC evolution from the literature and speculate on the ability of optical surveys to identify CCs. Finally, in \S5 the implications of these results and future avenues of research are summarized.

Throughout this Letter, we assume $H_0=$ 71 km s$^{-1}$ Mpc$^{-1}$, $\Omega_M=$ 0.73, and $\Omega_{vac}=$ 0.27.


\section{SDSS Data and Analysis}
In recent years, the number of known galaxy clusters in the range $0.1<z<0.6$ has increased significantly due to the application of complex cluster-finding algorithms to the wealth of publicly-available data in the Sloan Digital Sky Survey\footnote{\url{http://www.sdss.org}} \citep[hereafter SDSS; ][]{abazajian09}. In this study, we consider three recently-released catalogs, each of which identify galaxy clusters based on SDSS optical photometry, using different algorithms, as described in Table \ref{table:sample}. From each of these catalogs, the BCG positions were cross-referenced with the MPA-JHU\footnote{http://www.mpa-garching.mpg.de/SDSS/DR7/} release of SDSS DR7 spectrum measurements, to obtain H$\alpha$, H$\beta$, [\ion{O}{2}], and [\ion{O}{3}] emission line luminosities. All spectroscopic measurements from the MPA-JHU release have already been corrected for Galactic reddening, following \cite{odonnell94}.

\begin{table*}[tb]
\centering
{\small
\begin{tabular}{c r c l l}
\hline\hline
Sample & Number of & Redshift & Detection & Reference\\
Name & Galaxies & Range & Method & \\
\hline
Wen+09 & 39,716 & $0.05<z<0.6$ & $\bullet$ Friend-of-friend algorithm & \cite{wen09}\\
 & & & $\bullet$ $N\geq8$ & \\
  & \\
GMBCG & $>$55,000 & $0.1<z<0.55$ & $\bullet$ Red sequence matching & \cite{hao10}\\
 & \\
Szabo+11 & 69,173 & $0.045<z<0.78$ & $\bullet$ Adaptive matcher filter & \cite{szabo11}\\
& & & in position, magnitude, & \\
& & &  and redshift & \\
\hline
\end{tabular}
}
\caption{Summary of optically-selected galaxy cluster catalogs used in this study. Further explanation of the detection algorithms and sample properties can be found in the individual catalogue references.}
\label{table:sample}
\end{table*}

\begin{figure*}[htb]
\centering
\includegraphics[width=0.95\textwidth]{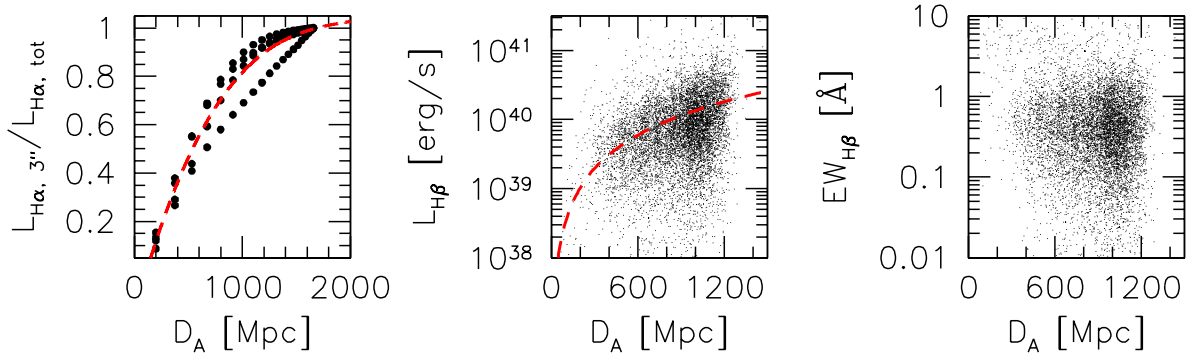}
\caption{Graphical depiction of three separate aperture corrections which have been used in this work. Left: Simulated fraction of H$\alpha$ luminosity within a 3$^{\prime\prime}$ aperture as a function of angular diameter distance, $D_A$, for systems with complex, extended H$\alpha$ filaments: Abell~0478, Abell~1795, Abell~2597, Sersic~159-03. The red, dashed line is a fit to these points and represents the first of three aperture corrections used in this work. Middle: H$_{\beta}$ luminosity as a function of $D_A$ for 12,230 BCGs in the GMBCG catalog. In order to correct for the increasing aperture size, we assume that the average H$_{\beta}$ luminosity is redshift-independent. The dashed red line represents the aperture correction required to remove the observed non-zero slope. Right: H$_{\beta}$ equivalent width, $EW_{H\beta}$ as a function of $D_A$. The use of $EW_{H\beta}$ assumes that the radial distribution of H$\beta$ luminosity is proportional to the continuum emission.}
\label{fig:apcorr}
\end{figure*}

In order to compare spectroscopic quantities over a range of redshift, a suitable aperture correction must be applied to account for the increased apparent diameter of the 3$^{\prime\prime}$ SDSS fiber aperture at larger distance. In this work, we consider three separate aperture corrections in an attempt to determine the \emph{total} emission-line luminosity from BCGs over the redshift range $0<z<0.6$. These three methods are:

\begin{enumerate}
\item {\bf Universal L$_{H\alpha}$(r) profile} -- Using deep, high spatial resolution H$\alpha$ imaging of four BCGs with the most luminous, extended, optically-emitting filaments from \cite{mcdonald10,mcdonald11a}, we derive a Universal L$_{H\alpha}$(r) profile, as shown in Figure \ref{fig:apcorr}a. This is used to estimate the fraction of the emission line flux which falls outside the 3$^{\prime\prime}$ aperture as a function of redshift (or apparent aperture size).
\item {\bf No evolution in $<L_{H\beta}>$} -- In Figure \ref{fig:apcorr}b the H$\beta$ luminosity is shown as a function of angular diameter distance for 12,230 BCGs in the GMBCG catalog. The trend towards higher emission-line luminosity at large distance is most likely a result of increasing aperture size. Thus, we make the assumption that the \emph{mean} H$\beta$ luminosity should be independent of redshift and simply fit a power law to these data. Applying this correction results in a flat distribution of L$_{H\beta}$ with distance.
\item {\bf L$_{H\alpha} \propto\ $L$_{continuum}$} -- By considering the equivalent width (line luminosity normalized to continuum luminosity), we make the assumption that the warm line-emitting gas has a similar radial distribution as the stellar content. This appears to be a reasonable assumption for an ensemble, as the equivalent width shows no strong evolution with angular diameter distance (Figure \ref{fig:apcorr}c).
\end{enumerate}

\begin{figure}[p]
\centering
\includegraphics[trim=1.0cm 1.5cm 1.0cm 1.5cm, width=0.6\textwidth]{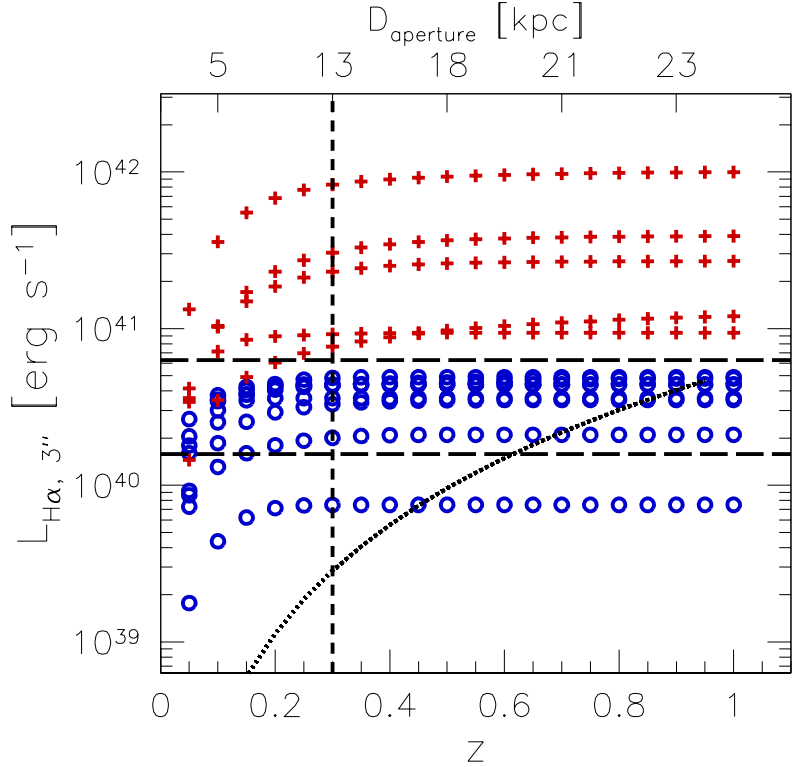}
\caption{Total H$\alpha$ luminosity within a 3$^{\prime\prime}$ diameter aperture versus redshift for a sample of 13 galaxy clusters from \cite{mcdonald10,mcdonald11a,mcdonald11b}.  Systems with complex, extended emission and nuclear emission  are depicted with red crosses and blue circles, respectively. Horizontal dashed lines represent luminosity limits of $10^{40.2}$ and $10^{40.8}$ erg s$^{-1}$, chosen to separate extended from compact emission and to allow detection of weak emitters at $z=0.6$. The curved dotted line represents the estimated sensitivity limits, showing that beyond $z\gtrsim0.4$, we may be missing a fraction of the weakly-emitting systems. For $z>0.3$,  the 3$^{\prime\prime}$ aperture encloses enough of the total H$\alpha$ luminosity to reliably separate these luminosity classes. The upper axis shows the angular diameter of a 3$^{\prime\prime}$ aperture for a given redshift.}
\label{fig:haflux}
\end{figure}

These three methods of estimating the total emission-line luminosity based on the central 3$^{\prime\prime}$ provide a range of aperture corrections which can then be compared. Both the magnitude and uncertainty of these aperture corrections is largest for nearby BCGs, where the 3$^{\prime\prime}$ aperture encloses only a small fraction of the total light.  In Figure \ref{fig:haflux}, the H$\alpha$ luminosity within a simulated 3$^{\prime\prime}$ aperture is shown over the range $0<z<1$ for 13 BCGs from \cite{mcdonald10,mcdonald11a}. This figure demonstrates that, for $z>0.3$, a sufficient fraction of the total H$\alpha$ luminosity is captured by a 3$^{\prime\prime}$ fiber to differentiate between strong (extended, complex filaments) and weak (nuclear, slightly extended) H$\alpha$ emission. Based on Figure \ref{fig:haflux}, we choose limiting H$\alpha$ luminosities of $10^{40.2}$ erg s$^{-1}$ and $10^{40.8}$ erg s$^{-1}$ to identify weak and strong optical emission

In order to increase signal-to-noise, emission from the H$\alpha$, H$\beta$, and [\ion{O}{2}] lines  were combined wherever possible. This is done assuming typical conversions of H$\alpha$/H$\beta$ = 2.85 \citep[case B recombination;][]{osterbrock89} and [\ion{O}{2}]/H$\beta$ = 1.4, which is typical of an \ion{H}{2} region. For each cluster, given a set of line fluxes and uncertainties, the probability that the optical emission lies above the aforementioned luminosity thresholds is computed assuming Gaussian statistics. The sum of these probabilities for all clusters within a given redshift bin then represents the fraction of clusters with optical line emission, N$_{OLE}$/N$_{total}$.  In order to account for contamination due to AGN, we remove all systems with [\ion{O}{3}]/H$\beta>3$, which is atypical for the optical filaments typically found in CC clusters \citep[][McDonald \etal in prep]{crawford99}.

\section{The Evolution of Optical Line-Emitting BCGs from $z=0$ to $z=0.6$}

\begin{figure*}[p]
\centering
\includegraphics[trim=2cm 1cm 2cm 1cm, width=0.85\textwidth]{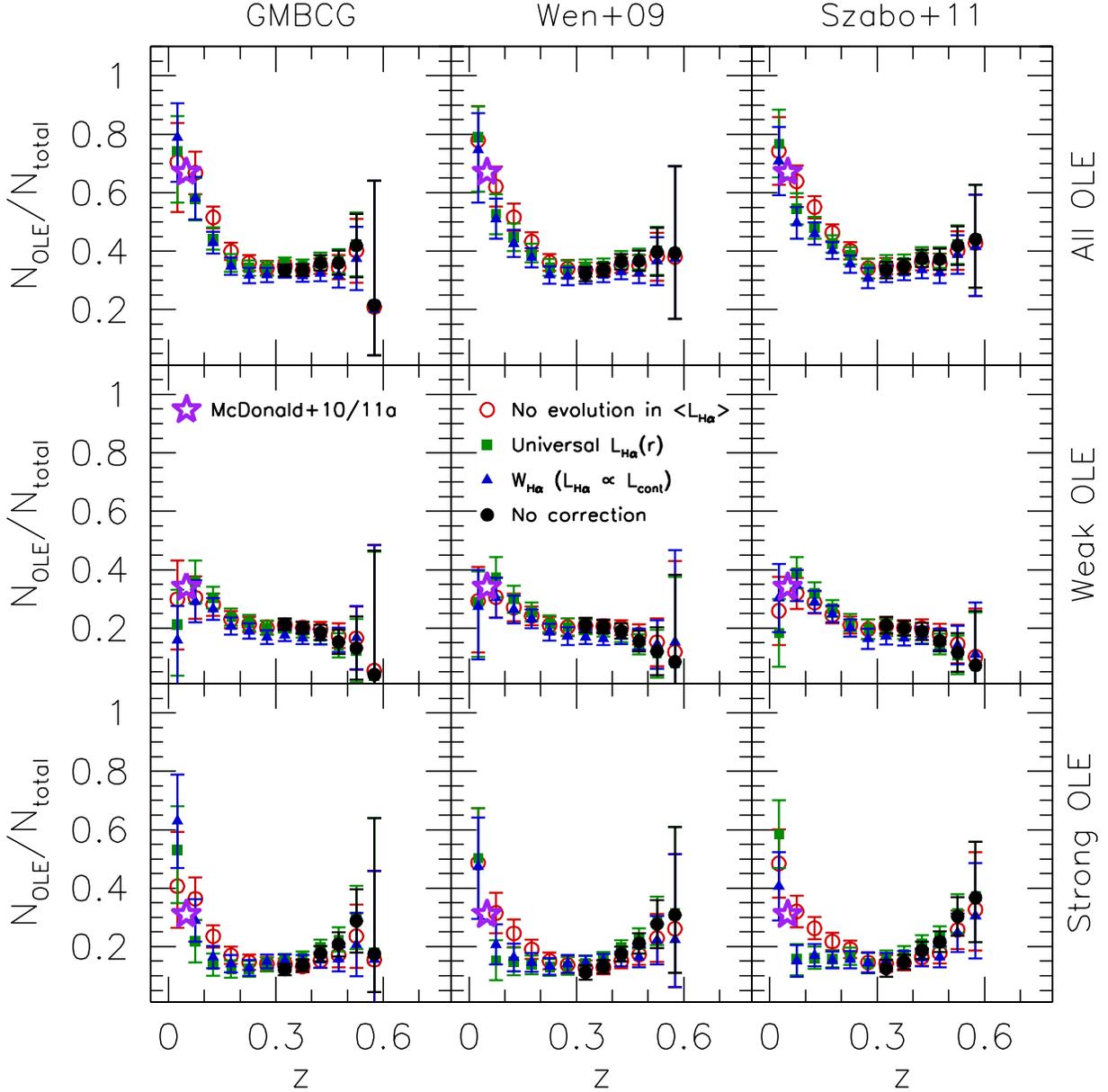}
\caption{Fraction of galaxy clusters with optical line emission (OLE) in the BCG as a function of redshift for three different optically-selected samples of galaxy clusters. The three samples, described in \S2, detect clusters using different techniques and yield $>10,000$ BCG spectra each. Point type and color represent the three aperture corrections described in \S2, while solid black points represent data for $z>0.3$ clusters with no aperture correction. The different catalogs and aperture corrections all yield similar results: there is a decrease in the number of galaxy clusters with emission line nebulae from $z=0$ to $z=0.3$, and an upturn in the fraction of systems with strong emission at $z>0.3$.}
\label{fig:ccfrac}
\end{figure*}

In Figure \ref{fig:ccfrac}, the fraction of galaxy clusters hosting an optical line-emitting BCG as a function of redshift is shown. For each of the three catalogs described in Table \ref{table:sample}, we show the fraction of clusters with optical line emission in the BCGs, as well as those with strong (L$_{H\alpha}>10^{40.8}$ erg s$^{-1}$) and weak ($10^{40.2}$ erg s$^{-1} <$ L$_{H\alpha}<10^{40.8}$ erg s$^{-1}$) emission separately. In general, there is little difference between the three cluster catalogs, despite the different cluster identification algorithms.  All nine panels of Figure \ref{fig:ccfrac} show the same overall trend, regardless of the emission strength, sample selection, or aperture correction used. The fraction of clusters with optical line-emitting BCGs decreases with redshift until it reaches a minimum at $z\sim0.3$. For $z>0.3$, the fraction of clusters with weak line emission continues to decrease, while the fraction with strong emission increases. The net effect of this opposite evolution is that the total fraction of line-emitting clusters remains constant over the range $0.3<z<0.6$. We caution that, while the line luminosity thresholds have been chosen to reduce sensitivity biases,  the fraction of clusters with weak optical emission may still be under-estimated at high redshift.

Figure \ref{fig:ccfrac} provides strong evidence for a minimum in the evolution of optical nebulae in galaxy clusters at $z\sim0.3$, or roughly 3.5 Gyr ago. The agreement between different indicators (H$\alpha$, H$\beta$, [\ion{O}{2}]), and different catalogs suggests that selection effects and measurement errors are minimal. What follows is a brief discussion of the two regimes bounding the minimum at $z\sim0.3$, and what processes may govern the evolution of optical nebulae at these epochs, followed by a summary of the different biases and systematic uncertainties which could conspire to produce the observed evolution.

\subsection{Optical Line-Emitting BCGs at $z<0.3$}
The evidence for a decline in the fraction of line-emitting BCGs over the past $\sim$3.5 Gyr is compelling. Interestingly, this timescale agrees with the finding by \cite{mcdonald10} that H$\alpha$ filaments extend to radii at which the ICM cooling time is $\sim$ 3-5 Gyr, \emph{but never beyond}. The presence of an apparent equilibrium between radio feedback from the central AGN and the cooling ICM \citep[e.g.,][]{sun09b} also suggests that cooling has been ongoing for a substantial amount of time. Combined, these results suggest that the majority of galaxy clusters have been relatively undisturbed over the past few Gyr, during which time the ICM has been allowed to cool, resulting in the formation of young stars, a multiphase ICM, and the establishment of a feedback cycle between the AGN and the cooling ICM.

One mechanism which could produce the observed evolution in Figure \ref{fig:ccfrac} over the range $0<z<0.3$ is cluster mergers. Based on simulations, \cite{genel09} predict that the fraction of massive halos undergoing major mergers drops by a factor of $\sim$4 between $z=1$ and $z=0$, while \cite{fakhouri10b} predict that 50\% of massive halos at $z=0$ had their last major merger at $z\geq$1.
Further, \cite{zuhone11} has shown that a merger between two galaxy clusters can fully disrupt a CC,  increasing the central cooling time to $t_{cool} \sim$ 10 Gyr. This would cut off the supply of cool gas, quenching star formation and resulting in a general lack of optical line emission on timescales of $\sim$100 Myr. The ability of major mergers to prevent ICM cooling, coupled with the rapid decrease in merger rates for massive halos at $z<1$ could produce the observed minimum in the fraction of galaxy clusters with emission-line nebulae at $z\sim0.3$. In the absence of strong mergers at $z<0.3$, clusters may cool unimpeded until the onset of AGN feedback, producing a significant reservoir of cool gas and, thus, a steep rise in the fraction of optically-emitting BCGs.

Strong AGN feedback is another way of reducing the amount of optical emission at $z\sim0.3$. During a strong outburst, the radio luminosity of the central AGN can surpass the $pdV$ work required to remove the CC \citep[e.g., ][]{sun09b, mcdonald11a}. This imbalance between feedback and cooling is seen in a small number of $z\sim0$ galaxy clusters, including Abell~2052 and Hydra A, which exhibit massive cavities in the ICM. If such outbursts were more common in the early Universe ($z>0.3$), then ICM cooling would be suppressed. Once the CC is destroyed, the AGN would lose its fuel supply and ICM cooling would proceed unimpeded, producing the cool gas required to generate significant emission line nebulae at $z<0.3$.
 
\subsection{Optical Line-Emitting BCGs at $z>0.3$}
In Figure \ref{fig:ccfrac}, we show that the fraction of BCGs with strong optical line emission increases with redshift beyond $z=0.3$, while the fraction with weak emission decreases over the same range. This decrease in the fraction of weak emitters may be a result of sensitivity limitations, despite efforts to set the threshold luminosity above than these limits (Figure \ref{fig:haflux}). The rise in the fraction of strong optical emitters is statistically significant at the 72\% and $>$99\% confidence levels for the GMBCG and Wen+09/Szabo+11 samples, respectively. This upturn suggests a significant amount of cool gas in clusters cores at $z\sim1$, roughly 6 Gyr after the Big Bang. This second maximum may correspond to the initial inflow of gas from the intergalactic medium onto the giant elliptical. Without a dense ICM at high redshift, mergers between galaxy groups will be less efficient in halting cooling. Indeed, \cite{evrard90} predicts that the majority of the ICM is formed around $z\sim1$ in a typical cluster and that most of this gas has $kT<1$ keV for $z\gtrsim1$. While these results are intriguing, deeper optical emission-line surveys are needed at $z>0.3$ in order to confirm the observed trends.

\subsection{Biases and Systematic Uncertainties}
The effects of various biases and systematic uncertainties were examined in order to assess their impact on the evolution presented in Figure \ref{fig:ccfrac}. BCG misidentification ($\sim$10\% in CCs; Leisman \etal in prep), bias towards rich clusters at high-z in optically-selected catalogs, and decreased sensitivity to H$\alpha$ at $z\sim0.3$ were found to have only a small impact on the observed evolutionary trends. As discussed previously, the evolution of optical line-emitting BCGs for $z<0.3$ is strongly dependent on the choice of aperture correction, but the fact that all three aperture corrections agree provides evidence that they are roughly correct. For $z>0.3$, the effects of aperture correction are negligible, but sensitivity biases may lead to an under-estimated fraction of weakly-emitting BCGs.


\section{Insights into the Evolution of Cool Cores?}

In recent years, the fraction of galaxy clusters hosting CCs has been quantified via a variety of methods, as reported in recent independent surveys by \cite{bauer05}, \cite{vikhlinin07}, \cite{ebeling07},  and \cite{santos08,santos10}. These studies are all based on X-ray data and rely on either the central cooling time or the ``cuspiness'' of the X-ray surface brightness profile to identify CCs.  While this is the most fundamental way of identifying a CC, which by definition has high density and short cooling times, it is also highly inefficient at larger redshift where the number of X-ray counts within the central $\sim$50 kpc is generally too low to properly constrain the central cooling time.  However, a more promising method may exist. In \cite{mcdonald10} we showed that the presence of H$\alpha$ filaments is evidence for cooling flows, with the H$\alpha$ luminosity being correlated (albeit with significant scatter) with both the X-ray cooling rate and central entropy. With this in mind, it is relevant to ask the question: Does the evolution of optical emission in galaxy clusters trace the evolution of CCs?

\begin{figure}[p]
\centering
\includegraphics[trim=2.0cm 1cm 13cm 1cm,width=0.3\textwidth]{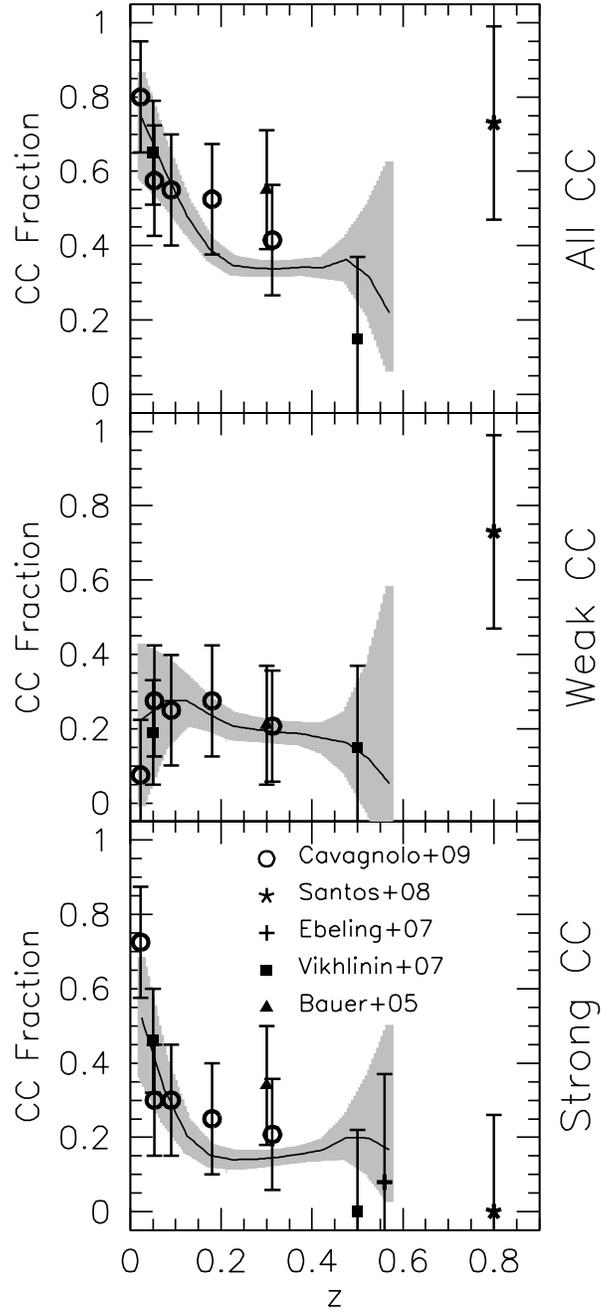}
\caption{The fraction of galaxy clusters with cool cores (CCs) as a function of redshift, based on the literature. The upper panel shows the general evolution of CCs, while the lower panels separate into weak and strong CCs. These classifications are based on central cooling time, X-ray surface brightness cuspiness, and central entropy. The shaded gray regions represent the evolution of optical line-emitting BCGs from Figure \ref{fig:ccfrac} (GMBCG catalog). The overall agreement between the CC and emission-line nebulae evolution suggests that large surveys for optical line emission in cluster cores may be able to accurately probe the CC evolution.}
\label{fig:literature}
\end{figure}

Figure \ref{fig:literature} shows the evolution of the CC fraction as a function of redshift. The four aforementioned X-ray studies provide estimates of both the weak and strong CC fraction at various epochs, but the data remain sparse.  We turn to the ACCEPT\footnote{\url{http://www.pa.msu.edu/astro/MC2/accept/}} database \citep{cavagnolo09} to add an additional 241 galaxy clusters with uniformly analyzed X-ray data. Using the readily-available central entropy ($K_0$) measurements from the ACCEPT database, the CC fraction is estimated over the range $0<z<0.4$ for weak ($20 < K_0 < 70$ keV cm$^2$), and strong ($K_0<20$ keV cm$^2$) CCs. This compilation represents the most detailed picture of the evolution of CCs in galaxy clusters to date. In \cite{mcdonald10} we showed that galaxy clusters with extended, filamentary H$\alpha$ emission also had the highest cooling rate and lowest central entropy. Thus, for a large enough sample, the evolution of strong optical line emission should mirror the evolution of strong CCs, and likewise for weak CCs. Overlaid on the X-ray data in Figure \ref{fig:literature}, we show the evolution of optical line-emitting BCGs from Figure \ref{fig:ccfrac}, for the GMBCG catalog. 
We find, overall, qualitative agreement between the evolution of optical nebulae and CCs in galaxy clusters, with the exception of any statistically significant increase in strong CCs at large redshift, providing additional evidence for a connection between the warm and hot ICM phase and justifying the use of optical line emission as a probe of CCs.

If, in an ensemble of clusters, the presence of optical line emission is indeed a tracer of CCs (as Figure \ref{fig:literature} seems to indicate), the discussion in \S3 can apply to CCs as well -- namely that there is a distinct lack of CCs 3.5 Gyr ago that may be due to increased merger or AGN activity in the early Universe.

\section{Summary}
The presence of optical emission-line nebulae in the cores of galaxy clusters is often taken as evidence for ongoing cooling of the intracluster medium (ICM). Thus, understanding the evolution of this warm phase is an important step towards understanding the heating/cooling balance in cluster cores over cosmic time. In this Letter, we exploit the availability of large, optically-selected catalogs of galaxy clusters with readily-available optical spectra in the Sloan Digital Sky Survey. We find evidence for a decline in the fraction of brightest cluster galaxies (BCGs) with optical emission-line nebulae from $z=0$ to $z=0.3$, suggesting that cooling has only recently begun to dominate feedback in cluster cores. For $z>0.3$, the fraction of clusters with strong optical emission rises, hinting at an earlier epoch of strong cooling at $z\sim1$. The evolution of optical emission in BCGs is compared to estimates of the cool core (CC) fraction and found to agree overall, with the exception that an upturn in the strong CC fraction is not observed. This agreement suggests that narrow-band optical surveys of high-$z$ galaxy clusters may provide a new, efficient way of probing the evolution of CCs at redshifts where X-ray methods become too expensive. In the future, deep, narrow-band H$\alpha$ imaging, which does not require complicated aperture corrections, of both low-z and high-z cluster samples will allow for an even more accurate analysis of high-$z$ emission-line nebulae.

\section*{Acknowledgements} 
MM was supported by NASA through SAO Award Number 2834-MIT-SAO-4018, issued by the Chandra X-ray Observatory Center on behalf of NASA (\#NAS8-03060). Thanks to M. Bautz, S. Courteau, R. Mushotzky, and S. Veilleux for useful comments and discussion on the original draft.


\begin{thebibliography}{29}
\expandafter\ifx\csname natexlab\endcsname\relax\def\natexlab#1{#1}\fi

\bibitem[{{Abazajian} {et~al.}(2009){Abazajian}, {Adelman-McCarthy},
  {Ag{\"u}eros}, {Allam}, {Allende Prieto}, {An}, {Anderson}, {Anderson},
  {Annis}, {Bahcall}, \& et~al.}]{abazajian09}
{Abazajian}, K.~N., {Adelman-McCarthy}, J.~K., {Ag{\"u}eros}, M.~A., {Allam},
  S.~S., {Allende Prieto}, C., {An}, D., {Anderson}, K.~S.~J., {Anderson},
  S.~F., {Annis}, J., {Bahcall}, N.~A., \& et~al. 2009, \apjs, 182, 543

\bibitem[{{Bauer} {et~al.}(2005){Bauer}, {Fabian}, {Sanders}, {Allen}, \&
  {Johnstone}}]{bauer05}
{Bauer}, F.~E., {Fabian}, A.~C., {Sanders}, J.~S., {Allen}, S.~W., \&
  {Johnstone}, R.~M. 2005, \mnras, 359, 1481

\bibitem[{{Cavagnolo} {et~al.}(2009){Cavagnolo}, {Donahue}, {Voit}, \&
  {Sun}}]{cavagnolo09}
{Cavagnolo}, K.~W., {Donahue}, M., {Voit}, G.~M., \& {Sun}, M. 2009, \apjs,
  182, 12

\bibitem[{{Churazov} {et~al.}(2001){Churazov}, {Br{\"u}ggen}, {Kaiser},
  {B{\"o}hringer}, \& {Forman}}]{churazov01}
{Churazov}, E., {Br{\"u}ggen}, M., {Kaiser}, C.~R., {B{\"o}hringer}, H., \&
  {Forman}, W. 2001, \apj, 554, 261

\bibitem[{{Conselice} {et~al.}(2001){Conselice}, {Gallagher}, \&
  {Wyse}}]{conselice01}
{Conselice}, C.~J., {Gallagher}, III, J.~S., \& {Wyse}, R.~F.~G. 2001, \aj,
  122, 2281

\bibitem[{{Cowie} {et~al.}(1980){Cowie}, {Fabian}, \& {Nulsen}}]{cowie80}
{Cowie}, L.~L., {Fabian}, A.~C., \& {Nulsen}, P.~E.~J. 1980, \mnras, 191, 399

\bibitem[{{Crawford} {et~al.}(1999){Crawford}, {Allen}, {Ebeling}, {Edge}, \&
  {Fabian}}]{crawford99}
{Crawford}, C.~S., {Allen}, S.~W., {Ebeling}, H., {Edge}, A.~C., \& {Fabian},
  A.~C. 1999, \mnras, 306, 857

\bibitem[{{Crawford} {et~al.}(2005){Crawford}, {Sanders}, \&
  {Fabian}}]{crawford05}
{Crawford}, C.~S., {Sanders}, J.~S., \& {Fabian}, A.~C. 2005, \mnras, 361, 17

\bibitem[{{Ebeling} {et~al.}(2007){Ebeling}, {Barrett}, {Donovan}, {Ma},
  {Edge}, \& {van Speybroeck}}]{ebeling07}
{Ebeling}, H., {Barrett}, E., {Donovan}, D., {Ma}, C.-J., {Edge}, A.~C., \&
  {van Speybroeck}, L. 2007, \apjl, 661, L33

\bibitem[{{Evrard}(1990)}]{evrard90}
{Evrard}, A.~E. 1990, \apj, 363, 349

\bibitem[{{Fabian}(1994)}]{fabian94}
{Fabian}, A.~C. 1994, \araa, 32, 277

\bibitem[{{Fakhouri} {et~al.}(2010){Fakhouri}, {Ma}, \&
  {Boylan-Kolchin}}]{fakhouri10b}
{Fakhouri}, O., {Ma}, C.-P., \& {Boylan-Kolchin}, M. 2010, \mnras, 406, 2267

\bibitem[{{Genel} {et~al.}(2009){Genel}, {Genzel}, {Bouch{\'e}}, {Naab}, \&
  {Sternberg}}]{genel09}
{Genel}, S., {Genzel}, R., {Bouch{\'e}}, N., {Naab}, T., \& {Sternberg}, A.
  2009, \apj, 701, 2002

\bibitem[{{Hao} {et~al.}(2010){Hao}, {McKay}, {Koester}, {Rykoff}, {Rozo},
  {Annis}, {Wechsler}, {Evrard}, {Siegel}, {Becker}, {Busha}, {Gerdes},
  {Johnston}, \& {Sheldon}}]{hao10}
{Hao}, J., {McKay}, T.~A., {Koester}, B.~P., {Rykoff}, E.~S., {Rozo}, E.,
  {Annis}, J., {Wechsler}, R.~H., {Evrard}, A., {Siegel}, S.~R., {Becker}, M.,
  {Busha}, M., {Gerdes}, D., {Johnston}, D.~E., \& {Sheldon}, E. 2010, \apjs,
  191, 254

\bibitem[{{Hatch} {et~al.}(2006){Hatch}, {Crawford}, {Johnstone}, \&
  {Fabian}}]{hatch06}
{Hatch}, N.~A., {Crawford}, C.~S., {Johnstone}, R.~M., \& {Fabian}, A.~C. 2006,
  \mnras, 367, 433

\bibitem[{{Heckman} {et~al.}(1989){Heckman}, {Baum}, {van Breugel}, \&
  {McCarthy}}]{heckman89}
{Heckman}, T.~M., {Baum}, S.~A., {van Breugel}, W.~J.~M., \& {McCarthy}, P.
  1989, \apj, 338, 48

\bibitem[{{McDonald} {et~al.}(2011{\natexlab{a}}){McDonald}, {Veilleux}, \&
  {Mushotzky}}]{mcdonald11a}
{McDonald}, M., {Veilleux}, S., \& {Mushotzky}, R. 2011{\natexlab{a}}, \apj,
  731, 33

\bibitem[{{McDonald} {et~al.}(2010){McDonald}, {Veilleux}, {Rupke}, \&
  {Mushotzky}}]{mcdonald10}
{McDonald}, M., {Veilleux}, S., {Rupke}, D.~S.~N., \& {Mushotzky}, R. 2010,
  \apj, 721, 1262

\bibitem[{{McDonald} {et~al.}(2011{\natexlab{b}}){McDonald}, {Veilleux},
  {Rupke}, {Mushotzky}, \& {Reynolds}}]{mcdonald11b}
{McDonald}, M., {Veilleux}, S., {Rupke}, D.~S.~N., {Mushotzky}, R., \&
  {Reynolds}, C. 2011{\natexlab{b}}, \apj, 734, 95

\bibitem[{{O'Donnell}(1994)}]{odonnell94}
{O'Donnell}, J.~E. 1994, \apj, 422, 158

\bibitem[{{Osterbrock}(1989)}]{osterbrock89}
{Osterbrock}, D.~E. 1989, {Astrophysics of gaseous nebulae and active galactic
  nuclei}

\bibitem[{{Santos} {et~al.}(2008){Santos}, {Rosati}, {Tozzi}, {B{\"o}hringer},
  {Ettori}, \& {Bignamini}}]{santos08}
{Santos}, J.~S., {Rosati}, P., {Tozzi}, P., {B{\"o}hringer}, H., {Ettori}, S.,
  \& {Bignamini}, A. 2008, \aap, 483, 35

\bibitem[{{Santos} {et~al.}(2010){Santos}, {Tozzi}, {Rosati}, \&
  {B{\"o}hringer}}]{santos10}
{Santos}, J.~S., {Tozzi}, P., {Rosati}, P., \& {B{\"o}hringer}, H. 2010, \aap,
  521, A64+

\bibitem[{{Sun}(2009)}]{sun09b}
{Sun}, M. 2009, \apj, 704, 1586

\bibitem[{{Szabo} {et~al.}(2011){Szabo}, {Pierpaoli}, {Dong}, {Pipino}, \&
  {Gunn}}]{szabo11}
{Szabo}, T., {Pierpaoli}, E., {Dong}, F., {Pipino}, A., \& {Gunn}, J. 2011,
  \apj, 736, 21

\bibitem[{{Vikhlinin} {et~al.}(2007){Vikhlinin}, {Burenin}, {Forman}, {Jones},
  {Hornstrup}, {Murray}, \& {Quintana}}]{vikhlinin07}
{Vikhlinin}, A., {Burenin}, R., {Forman}, W.~R., {Jones}, C., {Hornstrup}, A.,
  {Murray}, S.~S., \& {Quintana}, H. 2007, in Heating versus Cooling in
  Galaxies and Clusters of Galaxies, ed. {H.~B{\"o}hringer, G.~W.~Pratt,
  A.~Finoguenov, \& P.~Schuecker }, 48--+

\bibitem[{{Wen} {et~al.}(2009){Wen}, {Han}, \& {Liu}}]{wen09}
{Wen}, Z.~L., {Han}, J.~L., \& {Liu}, F.~S. 2009, \apjs, 183, 197

\bibitem[{{Werner} {et~al.}(2011){Werner}, {Sun}, {Bagchi}, {Allen}, {Taylor},
  {Sirothia}, {Simionescu}, {Million}, {Jacob}, \& {Donahue}}]{werner11}
{Werner}, N., {Sun}, M., {Bagchi}, J., {Allen}, S.~W., {Taylor}, G.~B.,
  {Sirothia}, S.~K., {Simionescu}, A., {Million}, E.~T., {Jacob}, J., \&
  {Donahue}, M. 2011, \mnras, 415, 3369

\bibitem[{{ZuHone}(2011)}]{zuhone11}
{ZuHone}, J.~A. 2011, \apj, 728, 54

\end{thebibliography}

\end{document}